\documentclass[10pt]{article} 
\usepackage[preprint]{tmlr}

\usepackage{amsmath,amsfonts,bm}

\def\eqref#1{equation~\ref{#1}}

\def\1{\bm{1}}

\DeclareMathAlphabet{\mathsfit}{\encodingdefault}{\sfdefault}{m}{sl}
\SetMathAlphabet{\mathsfit}{bold}{\encodingdefault}{\sfdefault}{bx}{n}

\usepackage{hyperref}
\usepackage{url}
\usepackage{amsmath}
\usepackage{physics}

\title{Biquaternion representation of the spin one half and it's application on the relativistic one electron atom. }

\author{\name Alejandro Arias Jiménez \email alejandro.ariasjz@udlap.mx \\
      \addr Department of Physics\\
      Universidad de las Américas Puebla
      \\
      }

\def\month{10}  
\def\year{2023} 
\def\openreview{\url{https://openreview.net/forum?id=XXXX}} 

\begin{document}

\maketitle

\begin{abstract}

In this work we represent the $1/2$ Spin particles with complex quaternions using a transformation to 2x2 matrices in order to obtain the Pauli matrices. With this representation we determine the states, rotation operators and the total angular momentum function in the complex quaternion space. Using this representation we work the solution for the relativistic hydrogen atom.

\end{abstract}

\section{Introduction.}

In quantum mechanics particles have a intrinsic angular momentum known as Spin, this properties changes between two types of particles, the Fermions and the Bosons. In specific, Fermions have a semi-integer spin and one of this type of particles are the electrons which have a $1/2$ spin, electrons are an elementary particle that conforms the Atom. (Sakurai, 1995). The Spin is a quantum phenomena that doesn't have an analog in classical physics.

The spin can be described as operators using Dirac formalism.

$$S_x \ket{\pm}=  \frac{\hbar}{2} \ket{-(\pm)} ,  $$
$$S_y \ket{\pm}= \pm i \frac{\hbar}{2} \ket{-(\pm)} , $$
$$ S_z \ket{\pm}= \pm \frac{\hbar}{2} \ket{\pm} . $$

But using the Heisenberg picture we can obtain a representation of the spin using the Pauli Matrices and the kets as $2 \times 1$ vectors and the bras as $1\times 2$ vectors.

$$S_x = \frac{\hbar}{2} \sigma_x $$
$$S_y = \frac{\hbar}{2} \sigma_y $$
$$S_z = \frac{\hbar}{2} \sigma_z $$
$$\ket{+}= \begin{pmatrix}
    1 \\
    0
\end{pmatrix} $$
$$\ket{-}= \begin{pmatrix}
    0 \\
    1
\end{pmatrix}$$

This matrices belong to the $SU(2)$ group and their product has similar properties to the quaternionic multiplication which lead us to think that we could obtain a quaternionic representation of the Spin $1/2$. 

This matrices are known as the Pauli matrices.

$$
\sigma_x=
\begin{pmatrix}
       0 & 1 \\
        1 &  0
    \end{pmatrix} 
,
\sigma_y=
\begin{pmatrix}
       0 & -i \\
        i &  0
    \end{pmatrix} 
,
\sigma_z=
\begin{pmatrix}
       1 & 0 \\
        0 &  -1
    \end{pmatrix} 
$$

\subsection{Objective.}

The main objective of this investigation is to obtain a complete complex quaternion representation of the spin $1/2$ particles and use it to solve quantum problems like the relativistic hydrogen atom.

The secondary objectives are: obtaining the representation of the spin rotation operator, applying quaternionic algebra to solve problems and applying the representation in different applications of the Pauli matrices. 
 
\subsection{Real quaternions.}

In order to obtain our representation we have to know what are quaternions. The quaternions are mathematical objets in $\mathbb{R}^4$ and are denoted by the space $\mathbb{H}(\mathbb{R})$ where the coefficients of each imaginary unit is real.

A quaternion is described in the following way.

\begin{equation}
    q= q_0 e_0 + q_1 e_1 + q_2 e_2 + q_3 e_3 
    \label{Eq1}
\end{equation}

This mathematical objects form a abelian group for the sum and a group that loses commutativity for the product. Following Froebenius Lem we now that objects from a 4rth dimension have a well defined product (Ward 1997).

The rule of the multiplication between the imaginary units of the quaternions is dictated by the Hamilton definition.

\begin{eqnarray*}
    \boldsymbol{e_1}\boldsymbol{e_2} = - \boldsymbol{e_2}\boldsymbol{e_1} = \boldsymbol{e_3}\\
    \boldsymbol{e_2}\boldsymbol{e_3} = - \boldsymbol{e_3}\boldsymbol{e_2} = \boldsymbol{e_1}\\
    \boldsymbol{e_3}\boldsymbol{e_1} = - \boldsymbol{e_1}\boldsymbol{e_3} = \boldsymbol{e_2}
\end{eqnarray*}

And product between two different quaternions is obtained the following way.

$$ q p = (a e_0 + b e_1 + c e_2 + d e_3) ( \alpha e_0 + \beta e_1 + \gamma e_2 + \tau e_3 ) \equiv $$ 
$$ a \alpha e_0 + a \beta e_1 + a \gamma e_2 + a \tau e_3 - b \alpha e_1 - b \beta e_0 + b \gamma e_3 - b \tau e_2 - c \alpha e_2 - c \beta e_3 - c \gamma e_0 + c \tau e_1 - d \alpha e_3 + d \beta e_2 - d \gamma e_1 - d \tau e_0 . $$

Simplifying by components. 

$$qp = (a \alpha - b \beta - c \gamma - d \tau) e_0 + $$ $$( a \beta - b \alpha + c \tau - d \gamma)e_1 + $$ $$(a \gamma - b \tau - c \alpha + d \beta) e_2 + $$ $$(a \tau + b \gamma - c \beta - d \alpha)e_3. $$

And defining the scalar part of the quaternion as $Sc(q)= q_0$ and the vectorial part as $Vec(q)= q_1 e_1 + q_2 e_2 + q_3 e_3$ we can obtain the following equation.

$$qp= Sc (q) Sc (p) - <q,p> + Sc(q) Vec(p) + Sc(p) Vec(q) + (q \cross p).$$

Another important aspect of the quaternionic algebra is the conjugate, defined similarly to the conjugate of a complex number.

$$\bar{q} =  q_0 e_0  - q_1 e_1 - q_2 e_2 - q_3 e_3 $$

With the conjugate the square norm of a quaternion is defined.

$$ |q|^2 = q \bar{q} = q_0 ^2 + q_1 ^2 + q_2 ^2 + q_3 ^2 = \sum_{i=0}^3 q_i ^2$$

Also doing a similar aproach to the complex numbers, the inverse of a quaternion can be defined in the following way.

$$q^{-1} = \frac{\bar{q}}{|q|^2}$$

Now that we know properties of the quaternions we can start with their representation in matrices. In order to obtain this, the quaternion is compacted into a complex number with complex numbers in their components.

$$q = z + we_2$$

Where

$$ q = z + we_2 =  (q_0 + q_1 e_1) + (q_2 + q_3 e_1 )e_2 = q_0 e_0 + q_1 e_1 + q_2 e_2 + q_3 e_3 . $$

Finally this compacted notation leads to the Matrix representation presented in work of Kravchenko and Shapiro. (Kravchenko V, Shapiro M, 1996).

\begin{equation}
Q=
    \begin{pmatrix}
        z & w \\
        -w^* & z^*
    \end{pmatrix}
\end{equation}

Real quaternions have interesting applications in physics, we can obtain a transformation to get the Pauli matrices in this space but for the hole spin representation we'll have to move on into the complex quaternions.

\subsection{Complex quaternions.}

Using equation \ref{Eq1} we can define the complex quaternions or biquaternions in a similar way to the real case, the only difference is that this time the coefficients of each imaginary unit is a complex number.

$$q = (a+ib)e_0 + (c + id) e_1 + (x +iy) e_2 + (g +ih) e_3 . $$

This mathematical objects are part of a $\mathbb{R}^8$ space and their imaginary unite commutes with the usual imaginary unit i. In the complex quaternions space there are certain objects called zero divisors but for our problem they become useful and we can treat them like the matrix zero divisors elements, it is important to mention that with these zero divisors the inverse quaternion is lost.

For the complex quaternions there are different types of conjugation we can do, the first one would be the conjugation of the vectorial part.

$$\bar{q} =  q_0 - q_1 - q_2 - q_3 = (a+ib)e_0 - (c + id) e_1 - (x +iy) e_2 - (g +ih) e_3 .$$

Another type of conjugation is conjugations each of the complex numbers.

$$ q^* =  q_0^* + q_1^* + q_2^* + q_3^* = (a-ib)e_0 + (c - id) e_1 + (x -iy) e_2 + (g -ih) e_3 .$$

Also, we can combine the other two conjugations in the same one.

$$ \bar{q}^* =  q_0^* - q_1^* - q_2^* - q_3^* = (a-ib) - (c - id) e_1 - (x -iy) e_2 - (g -ih) e_3. $$

With this conjugation we can define the square norm.

\begin{equation}
   |q|^2 = Sc(q\bar{q}^*) = \alpha^2 + \beta^2.
\end{equation}

Where $\alpha$ is.

$$\alpha ^2 = Re(q_0)^2 + Re(q_1)^2 + Re(q_2)^2 + Re(q_3)^2 . $$

And $\beta$.

$$\beta ^2= Im(q_0)^2 + Im(q_1)^2 + Im(q_2)^2 + Im(q_3)^2 .$$

The quadratic form that will be used as the norm for this work can be obtained in the following way.  

$$ Sc(q \bar{q}^*)= [(a+ib) + (c + id) e_1 + (x +iy) e_2 + (g +ih) e_3][(a-ib) - (c - id) e_1 - (x -iy) e_2 - (g -ih) e_3.] \equiv $$

$$a^2 +b^2 + c^2 + d^2 + x^2 + y^2 + g^2 +h^2 = a^2 + c^2 + x^2 + g^2 + b^2 + d^2 + y^2 + h^2 \equiv$$

$$ \alpha ^2 + \beta ^2 = Sc(q \bar{q}^*) = |q|^2 .$$

To obtain the matrix represetation a compact notation is defined as.

$$q= z + we_2 = (q_0 + q_1 e_1) + (q_2 + q_3 e_1)e_2$$

With z and w being bicomplex numbers. With this notation the matrix representation can be obtained.

$$ Q=
    \begin{pmatrix}
        z & w \\
        -w^* & z^* 
    \end{pmatrix}
$$

This matrix stated by Kravchenko and Shapiro can be used for several applications but for the interest of this work another transformation needs to the stated.

\subsection{Hydrogen atom.}

In quantum mechanics the Schrodinger equations is used to obtain the energy values and the wave function of a physics problem. The equation is defined as.

$$i\hbar \frac{\partial}{\partial t} \psi = \hat{H} \Psi$$

Where $\hat{H}$ denotes the quantum Hamiltonian of the problem and $\Psi$ denotes the wave function. 

If the potential of our problem is independent of time, then we can use separation of variables method and arrive to the time-independent Schrodinger equation.

\begin{equation}
    \hat{H}\psi = E \psi .
    \label{eq1}
\end{equation}

For the Hydrogen atom problem the Hamiltonian is defined in the following way. 

$$\hat{H} = \frac{-\hbar^2}{2m} \Delta  - \frac{e^2}{4 \pi \epsilon_0 r} $$

Where $\Delta$ is the Laplace operator, $m$ is the mass of the particle, $\hbar^2$ is the reduced Plank's constant and $\frac{e^2}{4 \pi \epsilon_0 r}$ is the Coulomb potential. 

In order to appropriately solve the equation, the Laplace operator needs to be written in spherical coordinates due to the nature of the problem.

$$\Delta_s = \frac{1}{r^2} (\frac{\partial}{\partial r} (r^2 \frac{\partial}{\partial r} )) + \frac{1}{r^2 sin(\theta)}  (\frac{\partial}{\partial \theta} (sin(\theta) \frac{\partial}{\partial \theta} ) ) + \frac{1}{r^2 sin^2(\theta)} \frac{\partial^2}{\partial \phi^2}   $$

Introducing this expression into the Hamiltonian we finally obtain the Schrodinger equation for the Hydrogen atom.

$$(\frac{-\hbar^2}{2m}(\frac{1}{r^2} (\frac{\partial}{\partial r} (r^2 \frac{\partial}{\partial r} )) + \frac{1}{r^2 sin(\theta)}  (\frac{\partial}{\partial \theta} (sin(\theta) \frac{\partial}{\partial \theta} ) ) + \frac{1}{r^2 sin^2(\theta)} \frac{\partial^2}{\partial \phi^2}) - \frac{e^2}{4 \pi \epsilon_0 r}) \psi = E \psi$$

The solutions to this equation can be obtained using variable separations or ladder operators. In both ways the wave function is described by the harmonic oscillators for the angular parts and a exponential combined with the Laguerre Polynomials for the radial component. 

$$\psi(r,\theta,\phi)= A e^{- \frac{r}{2}} r^l L_{n+l} ^{2l+1} (r) Y_l ^m (\theta, \phi) $$

This solution can be used to formalize atomic and molecular physics but it is incomplete due to the relativistic nature of the electron and the spin orbit correction. Then, in order to achieve this solution we need to include more terms to the Hamiltonian. 

$$ \hat{H} = \frac{\Vec{p}^2}{2m}  + \frac{e^2}{r} -\frac{\Vec{p}^4}{8m^3 c^2} + \frac{1}{2m^2 c^2} \frac{1}{r} \frac{dV}{dr} \Vec{S} \cdot \Vec{L} + \frac{\hbar^2}{8m^2 c^2}(4\pi e^2 \delta(\Vec{r})) $$

This Hamiltonian leads to the corrections in the energy values using perturbation theory. But to obtain the eigenfunctions we need to use the Dirac Hamiltonian and substitute in \ref{eq1}.

$$ \hat{H}= \alpha \cdot \hat{p} + \beta m - \frac{Z e^2}{r}  $$

Finally arriving to the Dirac equation for the relativistic one electron atom.

\begin{equation}
    \Vec{\alpha} \cdot \Vec{p} \Psi + \beta mc^2 \Psi -\hbar c \frac{z a_0}{r} \Psi = E\Psi
    \label{DiracE}
\end{equation}

\subsection{Relativistic one electron atom.}

In order to solve \ref{DiracE} for the problem we need to change Cartesian's coordinates to spherical. Where $\Sigma$ are the total Spin matrices.

$$ (\alpha_r (p_r + \frac{i}{r} (\Vec{\Sigma} \cdot \Vec{L} + \hbar))) \Psi + \beta mc^2 \Psi -\hbar c \frac{z a_0}{r} \Psi = E\Psi $$

Now we arrive into a separable partial differential equation where the total function can be divided into $\Psi(r,\theta , \phi) = R(r)\Theta(\theta,\phi)$.

For the angular part we arrive at the following equation. 

$$(\Vec{\Sigma} \cdot \Vec{L} + \hbar)\Psi = \frac{1}{\hbar} (\Vec{J}^2 + \frac{\hbar^2}{4} - \Vec{L}^2) \Psi = \hbar (j(j+1) +\frac{1}{4} - l(l+1)) \Psi$$

The solution to this equation should be a function of the total angular momentum, also known as a Spinor function. 

$$
y_l ^{l \pm \frac{1}{2}}=
\frac{1}{\sqrt{2l+1}}
\begin{pmatrix}
     \pm \sqrt{l \pm m_j + \frac{1}{2}} Y_l ^{m_j-\frac{1}{2}} \\
      \sqrt{l - \pm m_j + \frac{1}{2}} Y_l ^{m_j+\frac{1}{2}}
\end{pmatrix}
$$

For the radial part we have the following equation.

\begin{equation}
    (\alpha _r p_r + i  (\alpha_r \beta \frac{c\hbar k}{r}) +\beta mc^2 -\hbar c \frac{Za_0}{r})R(r) = ER(r)
    \label{radio}
\end{equation}

Where $R(r)= \frac{1}{r}\begin{pmatrix}
        F(r)  \\
        G(r)
    \end{pmatrix} $ and $\alpha_r = \sigma_y$ ,$\beta = \sigma_z$, $p_r= -i\hbar \frac{1}{r} \frac{\partial}{\partial r} (r())$. 

Using these conditions we arrive to the following expression. 

$$\begin{pmatrix}
        -c \hbar \frac{dG}{dr} -\frac{c \hbar k}{r} G +mc^2 F - \frac{z\hbar c a_0}{r}F - E F  \\
        c \hbar \frac{dF}{dr} -\frac{c \hbar k}{r} F -mc^2 G - \frac{z\hbar c a_0}{r}G -EG
    \end{pmatrix}  = 0 $$

Solving each equation and multiplying by the angular part we obtain the solution of the problem.

\begin{equation}
    \Psi(r,\theta,\phi) = \frac{A}{r} \begin{pmatrix}
        F(\rho) y^{jm}_{j-\frac{1}{2}} \\
        i G(\rho)y^{jm}_{j+\frac{1}{2}}
    \end{pmatrix}
    \label{Sol}
\end{equation}

Where $F(\rho)$ is defined as.

$$F(\rho)= \rho ^s e^{-\rho} ((s-k) 2\rho L_{n-|k|-1} ^{2s+1} (2\rho) +z a_o \frac{smc^2 -kE}{\hbar c C}) L_{n-|k|} ^{2s-1} (2\rho) $$

While $G(\rho)$ is defined as.

$$G(\rho)= \rho ^s e^{-\rho} ( z a_o 2\rho L_{n-|k|-1} ^{2s+1} (2\rho) + (s-k)\frac{smc^2 -kE}{\hbar c C}) L_{n-|k|} ^{2s-1} (2\rho) $$

And 

$$ \rho = C r, C= \frac{\sqrt{m^2 c^4 - E^2}}{\hbar c}, s=\sqrt{k^2 - z^2 a_o^2}$$

This solution is obtained with a matrix representation and the total solution being a vector of 4 components. The eigenvalues or energies of this problem are given by.

$$E= mc^2 (1 + \frac{za_0}{n-|k| + \sqrt{k^2 - Z^2 a_0 ^2}})^{-\frac{1}{2}} $$

Now we need to begin the complex quaternion representation to obtain the solution in a different way.

\section{Complex quaternion representation of the spin 1/2.}

The complex quaternions transformation to matrices is proposed.

\begin{equation}
    A(q)= \begin{pmatrix}
       q_0 + i q_1 & q_2 + i q_3 \\
        q_2^* + (iq_3)^* &  q_0 - (iq_1)^* 
    \end{pmatrix}
\label{Transfomracion de Alex}
\end{equation}

Using unitary quaternions we arrive to the following expressions.

$$q=1e_0 + 0e_1 + 0e_2 + 0e_3 \longrightarrow A(q)= \begin{pmatrix}
       1 & 0 \\
        0 &  1
    \end{pmatrix} 
$$

$$q=0e_0 + -ie_1 + 0e_2 + 0e_3 \longrightarrow A(q)= \begin{pmatrix}
       1 & 0 \\
        0 &  -1
    \end{pmatrix} 
$$

$$q=0e_0 + 0e_1 + -ie_2 + 0e_3 \longrightarrow A(q)= \begin{pmatrix}
       0 & -i \\
        i &  0
    \end{pmatrix} 
$$

$$q=0e_0 + 0e_1 + 0e_2 - ie_3 \longrightarrow A(q)= \begin{pmatrix}
       0 & 1 \\
        1 &  0
    \end{pmatrix} 
$$

Leaving us with the representation of the Pauli matrices. 

$$I = q_I = 1e_0 +0e_1 + 0e_2 +0e_3 $$

$$\sigma_x = q_x = 0e_0 +0e_1 + 0e_2 - ie_3$$

$$\sigma_y = q_x = 0e_0 +0e_1 -ie_2 +0e_3$$

$$\sigma_z = q_z = 0e_0 -ie_1 + 0e_2 +0e_3$$

With all these equations we can propose the Spin operator representation.

\begin{equation}
Sx = \frac{\hbar}{2} q_x = \frac{\hbar}{2} (-ie_3)
\end{equation}

\begin{equation}
    Sy = \frac{\hbar}{2} q_y = \frac{\hbar}{2} (-ie_2) 
\end{equation}

\begin{equation}
    Sz = \frac{\hbar}{2} q_z = \frac{\hbar}{2} (-ie_1)
\end{equation}

And for the states the following representation for a $2 \times 1$ vector is proposed

$$ a(q)= \frac{\sqrt{2}}{2}
\begin{pmatrix}
    q_0 + i q_1 \\
    -q_2 + i q_3
    \label{a+}
\end{pmatrix}$$

Similarly the $1x2$ vector transformation is proposed.

$$ a^- (q)= \frac{\sqrt{2}}{2}
\begin{pmatrix}
    q_0 + i q_1 &
    q_2 - i q_3
    \label{a-}
\end{pmatrix}$$

With this transformation we arrive to the states representation.

\begin{equation}
    \ket{+} = q_+ = \frac{1}{\sqrt{2}} (e_0 - i e_1)
\end{equation}

\begin{equation}
    \ket{-} = q_- = \frac{1}{\sqrt{2}} (-e_2 - i e_3)
\end{equation}

To prove the representation we need to verify the eigenvalue equations for each spin operator.
Beginning with the spin in the X direction.

$$ \frac{\hbar}{2} q_x q_+ = \frac{\hbar}{2} \frac{1}{\sqrt{2}} (-ie_3)(e_0 -ie_1)= \frac{\hbar}{2} \frac{1}{\sqrt{2}} (-ie_3 -e_2) = \frac{\hbar}{2} q_- $$

Now for the down state.

$$\frac{\hbar}{2} q_x q_- = \frac{\hbar}{2} \frac{1}{\sqrt{2}} (-ie_3)(-e_2 -ie_3)= \frac{\hbar}{2} \frac{1}{\sqrt{2}} (-ie_1 + e_0) = \frac{\hbar}{2} q_+- $$

Now proving the spin in Y direction.

$$ \frac{\hbar}{2} q_y q_+ = \frac{\hbar}{2} \frac{1}{\sqrt{2}} (-ie_2)(e_0 -ie_1)= \frac{\hbar}{2} \frac{1}{\sqrt{2}} (-i e_2 + e_3) = i \frac{\hbar}{2} q_- $$

For the down state.

$$\frac{\hbar}{2} q_y q_- = \frac{\hbar}{2} \frac{1}{\sqrt{2}} (-ie_2)(-e_2-ie_3)= \frac{\hbar}{2} \frac{1}{\sqrt{2}} (-ie_0 -e_1) = -i \frac{\hbar}{2} q_+ $$

Finally the spin in the Z direction.

$$ \frac{\hbar}{2} q_z q_+ = \frac{\hbar}{2} \frac{1}{\sqrt{2}} (-ie_1)(e_0 -ie_1)= \frac{\hbar}{2} \frac{1}{\sqrt{2}} (-ie_1 +e_0) = \frac{\hbar}{2} q_+ $$

With the down state.

$$ \frac{\hbar}{2} q_z q_- = \frac{\hbar}{2} \frac{1}{\sqrt{2}} (-ie_1)(-e_2 -ie_3)= \frac{\hbar}{2} \frac{1}{\sqrt{2}} (ie_3 + e_2) = - \frac{\hbar}{2} q_- $$

And proving the Dirac formalism definition for $S_z$.

$$\frac{2}{\hbar} S_z = (\ket{+} \bra{+} - \ket{+} \bra{-}) = \frac{1}{4} ((e_0 -ie_1)(e_0 -ie_1) - (-e_2 -ie_3)(e_2 -ie_3)) = \frac{1}{4}(2e_0 -2ie_1 -2e_0 -2ie_1) = (-ie_1) $$

For $S_x$.

$$\frac{2}{\hbar} S_x = (\ket{+} \bra{-} + \ket{-} \bra{+}) = \frac{1}{4} ((e_0 -ie_1)(e_2 -ie_3) + (-e_2 -ie_3)(e_0 -ie_1)) = \frac{1}{4}(2e_2 -2ie_3 -2e_2 -2ie_3) = (-ie_3) $$

And for $S_y$.

$$\frac{2}{\hbar} S_y = -i (\ket{+} \bra{-} - \ket{-} \bra{+}) = -i \frac{1}{4} ((e_0 -ie_1)(e_2 -ie_3) - (-e_2 -ie_3)(e_0 -ie_1)) = -i \frac{1}{4}(2e_2 -2ie_3 +2e_2 +2ie_3) = (-ie_2) $$

Proving these eigenvalue equations we lastly need to check the orthogonality and the normalization of the states. 

For the norm we obtain the following expression.

$$ |q_+|^2 =\frac{1}{2}((1)^2 + (-1)^2) = 1$$

$$ |q_-|^2 =\frac{1}{2}((-1)^2 + (-1)^2) = 1$$

Now, to prove the orthogonality we define the dual states with the transformation \ref{a-}.

$$\bra{+}= \frac{1}{\sqrt{2}} (e_0 -ie_1) = \bar{q_+}^*$$

$$\bra{-}= \frac{1}{\sqrt{2}} (e_2 -ie_3) = \bar{q_-}^*$$

Arriving to the orthogonality.

$$\bar{q_+}^* q_- =\frac{1}{2} (e_0 -ie_1) (-e_2 - ie_3) = 0$$

$$\bar{q_-}^* q_+ =\frac{1}{2} (e_2 -ie_3) (e_0 -ie_1) = 0$$

Defined by the norm we proposed in the complex quaternion section.

The complex quaternion representation of the $1/2$ Spin is completed and now we can focus on the applications.

\section{Rotation Operator.}

In quantum mechanics the rotation operator using the spin is defined by this expression.

$$\hat{D}(n,\phi)= I\cos(\frac{\phi}{2})-i(\sigma \cdot n) \sin(\frac{\phi}{2})$$

And the hermitian conjugate defined as.

$$\hat{D}^{\dag} (n,\phi)= I \cos(\frac{\phi}{2})+i(\sigma \cdot n) \sin(\frac{\phi}{2})$$

Then, using our representation we can obtain the following quaternion representation.

$$\hat{D}(n,\phi)= e_o \cos(\frac{\phi}{2})-i(q_p \cdot n) \sin(\frac{\phi}{2})$$

Leading to the different rotations among the x,y and z axis. 

\begin{equation}
    \hat{D}(x,\phi)= e_o \cos(\frac{\phi}{2})-i(-ie_3) \sin(\frac{\phi}{2}) = e_o \cos(\frac{\phi}{2})- e_3 \sin(\frac{\phi}{2})
\end{equation}

\begin{equation}
    \hat{D}(y,\phi)= e_o \cos(\frac{\phi}{2})-i(-ie_2) \sin(\frac{\phi}{2}) = e_o \cos(\frac{\phi}{2})-e_2  \sin(\frac{\phi}{2})
\end{equation}
    
\begin{equation}
    \hat{D}(z,\phi)= e_o \cos(\frac{\phi}{2})-i(-ie_1) \sin(\frac{\phi}{2}) = e_o \cos(\frac{\phi}{2})-e_1 \sin(\frac{\phi}{2}) 
\end{equation}

And rotating the spin operator we obtain the same expression as in the matrix representation.

\begin{equation}
    D^ \dagger (n_i, \phi) S_j D(n_i, \phi) = \frac{\hbar}{2} (q_j \cos(\phi) + q_k \sin(\phi) )
\end{equation}

With these expressions we have rotations using complex quaternion representations.

\section{Raising and lowering operators representation.}

Another important operator is the raising and lowering operator also known as ladder operators for the spin, defined by the Spin in x and y direction.

$$S_+ = \frac{1}{2} (S_x + i S_y) $$

$$S_- = \frac{1}{2} (S_x - i S_y)$$

And using our transformation to the complex quaternions space we obtain the following operators.

\begin{equation}
    q^+ = \frac{1}{2} (e_2 - ie_3)
\end{equation}

\begin{equation}
    q^- = \frac{1}{2} (-e_2 - ie_3)
\end{equation}

Which means that we can obtain each operator by taking the complex quaternionic conjugate of the operator. This operators gives us a way to move between the up and down spin.

\section{Spinor functions.}

In quantum mechanics the total angular momentum functions are defined in the following way. 

$$
y_l ^{l \pm \frac{1}{2}}=
\frac{1}{\sqrt{2}}
\begin{pmatrix}
     \pm \sqrt{l \pm m_j + \frac{1}{2}} Y_l ^{m_j-\frac{1}{2}} \\
      \sqrt{l - \pm m_j + \frac{1}{2}} Y_l ^{m_j+\frac{1}{2}}
\end{pmatrix}
$$

Transforming it into the complex quaternion representations we obtain the following expression where $C_{l1} = \sqrt{l \pm m_j + \frac{1}{2}} $ and $C_{l2} = \sqrt{l - \pm m_j + \frac{1}{2}} $.

\begin{equation}
    y_l ^{l \pm \frac{1}{2}}= \frac{1}{\sqrt{2}} (C_{l1} Y_l ^{m_j-\frac{1}{2}} e_o -i C_{l1} Y_l ^{m_j-\frac{1}{2}} e_1 - C_{l2} Y_l ^{m_j+\frac{1}{2}} e_2 - iC_{l2} Y_l ^{m_j+\frac{1}{2}} e_3)        
\end{equation}

This function works in the same way as the matrix representation, we can compare them with a problem where we want to measure the probability to find the state down in a experiment given by the function. 

$$y_2 ^{2 \pm \frac{1}{2}} = \begin{pmatrix}
     \frac{2}{\sqrt{5}} Y_2 ^1 \\
      \frac{1}{\sqrt{5}} Y_2 ^2
\end{pmatrix} $$

Using the complex quaternion representation we arrive to this expression. 

$$ y_2 ^{2 \pm \frac{1}{2}} = \frac{2}{\sqrt{5}} Y_2 ^1 e_0 - \frac{2i}{\sqrt{5}} e_1 - \frac{1}{\sqrt{5}} Y_2 ^2 e_2 - \frac{i}{\sqrt{5}} Y_2 ^2 e_3 $$

And the probability to measure the down state in this total angular momentum function can be calculated in the complex quaternion representation.

$$ |\bar{q_-}^*  y_2 ^{2 \pm \frac{1}{2}} |^2  =  \frac{1}{\sqrt{2}} ((e_2 -ie_3)(\frac{2}{\sqrt{5}} Y_2 ^1 e_0 - \frac{2i}{\sqrt{5}} e_1 - \frac{1}{\sqrt{5}} Y_2 ^2 e_2 - \frac{i}{\sqrt{5}} Y_2 ^2 e_3))$$

Doing the calculations we arrive to the following expression.

$$ | \frac{1}{2} (\frac{2}{\sqrt{5}} Y_2 ^1 e_2 + \frac{2i}{\sqrt{5}} e_3 + \frac{1}{\sqrt{5}} Y_2 ^2 e_0  - \frac{i}{\sqrt{5}} Y_2 ^2 e_1 - \frac{2i}{\sqrt{5}}  Y_2 ^1 e_3 - \frac{2}{\sqrt{5}} Y_2 ^1 e_2 + \frac{2i}{\sqrt{5}} Y_2 ^1 e_3 + \frac{1}{\sqrt{5}} Y_2 ^2 e_0 ) | = | \frac{1}{2} \frac{2}{\sqrt{5}} Y_2 ^2 |^2 $$

And simplifying this result.

$$ |\bar{q_-}^*  y_2 ^{2 \pm \frac{1}{2}} |^2 = \frac{1}{5} |Y_2 ^2|^2 $$

Finally arriving to the same result we obtain using matrix representation or brackets formalism. In order to apply this representation we include one problem where the total angular momentum function is used and use the proposed complex quaternion formalism to solve it. 

The problem proposed is the solution of the relativistic hydrogen atom.

\section{Relativistic one electron atom solution using the complex quaternion representation.}

One application of the Spin and the Spinor functions is the solution of the relativistic one electron atom. In this case we already know that the angular solution is given by the Spinor functions which we can represent with complex quaternions. But there is also another interesting use, the solution of the radial equation \ref{radio}. This equation uses the Pauli Matrices in the y and z axis which can be substitute by our representation $\alpha_r = -ie_2, \beta=-ie$ converting it into the following equation.

\begin{equation}
    ((-ie_2) -i\hbar \frac{1}{r} \frac{\partial}{\partial r} (r()) + i(-ie_2)(-ie_1) \frac{c\hbar k}{r} + (-ie_1) mc^2 -\hbar c \frac{Za_0}{r})R(r) = ER(r)
\end{equation}

Where the $R(r)$ is given by a quaternion $R(r)= \frac{1}{r} (F(r)(e_o-ie_1) + G(r)(-e_2 -ie_3)) $. Using this definition we arrive to the radial quaternionic equation. 

\begin{eqnarray}
    -c \hbar \frac{dG}{dr}(e0- ie_1) -\frac{c \hbar k}{r} G (e_0- ie_i) + mc^2 F(e_0- ie_1) - \frac{z\hbar c a_0}{r}F(e_0 -ie_1) + 
    c \hbar \frac{dF}{dr} (-e_2 -ie_3) -\nonumber \\ \frac{c \hbar k}{r} F(-e_2 -ie_3) -mc^2 G (-e_2 -ie_3) - \frac{z\hbar c a_0}{r}G (-e_2 -ie_3) = EF(e_0 -ie_1) + EG(-e_2 -ie_3) 
\end{eqnarray}

This equation can be separated into a system of two differential equations same to the matrix case. 

\begin{equation}
    -c \hbar \frac{dG}{dr} -\frac{c \hbar k}{r} G  + mc^2 F - \frac{z\hbar c a_0}{r}F = EF
\end{equation}

\begin{equation}
    c \hbar \frac{dF}{dr} - \frac{c \hbar k}{r} F -mc^2 G  - \frac{z\hbar c a_0}{r}G = EG 
\end{equation}

Finally, the solution of the problem can be seen in two different ways using the complex quaternion representation of the spinor function and leaving it in a vector of two components.

\begin{equation}
    \Psi(r,\theta,\phi) = \frac{A}{r} \begin{pmatrix}
        F(\rho) y^{jm}_{j-\frac{1}{2}} \\
        i G(\rho)y^{jm}_{j+\frac{1}{2}}
    \end{pmatrix} =  \frac{A}{r} \begin{pmatrix}
        F(\rho) \frac{1}{\sqrt{2}} (C_{1} Y_k ^{m_j-\frac{1}{2}} e_o -i C_{1} Y_k ^{m_j-\frac{1}{2}} e_1 - C_{2} Y_k ^{m_j+\frac{1}{2}} e_2 - iC_{2} Y_k ^{m_j+\frac{1}{2}} e_3)   \\
        i G(\rho) \frac{1}{\sqrt{2}} (C_{3} Y_{-k} ^{m_j-\frac{1}{2}} e_o -i C_{3} Y_{-k} ^{m_j-\frac{1}{2}} e_1 - C_{4} Y_{-k} ^{m_j+\frac{1}{2}} e_2 - iC_{4} Y_{-k} ^{m_j+\frac{1}{2}} e_3)  
    \end{pmatrix}
\end{equation}

Where the $C_{n}$ denotes the Clebsch-Jordan coefficient of that particular case. Or we can arrive to a quaternion representation.

\begin{eqnarray}
    \Psi =\frac{1}{\sqrt{2}r} ( C_{1} F Y_k ^{m_j-\frac{1}{2}} +i C_{3} G Y_{-k} ^{m_j-\frac{1}{2}} )e_0 + (-iC_{1} F Y_k ^{m_j-\frac{1}{2}} + C_{3}G Y_{-k} ^{m_j-\frac{1}{2}}) e_1 \nonumber \\+ (-C_{2}F Y_k ^{m_j+\frac{1}{2}} + i C_{4}G Y_{-k} ^{m_j+\frac{1}{2}})e_2 + (-iC_{2}F Y_k ^{m_j +\frac{1}{2}} +C_{4} G Y_{-k} ^{m_j +\frac{1}{2}})e_3
\end{eqnarray}

With this representation the problem is solved leaving us with the wave function of the relativistic one electron atom in the complex quaternion space.

The dual state can be defined with the complex conjugate wave function.

\begin{eqnarray}
    \bar{\Psi^*} =\frac{1}{\sqrt{2}r} ( C_{1} F Y_k ^{m_j-\frac{1}{2}} -i C_{3} G Y_{-k} ^{m_j-\frac{1}{2}} )e_0 - (iC_{1} F Y_k ^{m_j-\frac{1}{2}} + C_{3}G Y_{-k} ^{m_j-\frac{1}{2}}) e_1 \nonumber \\- (-C_{2}F Y_k ^{m_j+\frac{1}{2}} - i C_{4}G Y_{-k} ^{m_j+\frac{1}{2}})e_2 - (iC_{2}F Y_k ^{m_j +\frac{1}{2}} +C_{4} G Y_{-k} ^{m_j +\frac{1}{2}})e_3
\end{eqnarray}

Arriving to the density probability function but only including the scalar term because that is what we need in order to obtain the norm,  the total probability density can be obtained using the quaternion multiplication rule.

\begin{eqnarray}
    Sc(\bar{\Psi^*} \Psi) = \frac{1}{2r^2}( 2(C_1 ^2 {Y_k ^{m_j-\frac{1}{2}}}^2 F^2 + C_2 ^2 {Y_k ^{m_j +\frac{1}{2}}} ^2 F^2 + C_3 ^2 {Y_{-k} ^{m_j-\frac{1}{2}}}^2  + C_4 ^2 {Y_{-k} ^{m_j +\frac{1}{2}}}^2 G^2 ) e_0 
\end{eqnarray}

Where $Y^2= Y^* Y$.

Using the proposed squared norm $ |\Psi|^2 = Sc(\bar{\Psi^*} \Psi )$ leaving us with only the scalar terms, we obtain the probability of finding the electron between a and -a where this limits are volume units.

\begin{equation}
    \rho = \int_{-a}^{a} \frac{1}{r^2} (C_1 ^2 {Y_k ^{m_j-\frac{1}{2}}}^2 F(r) ^2 + C_2 ^2 {Y_k ^{m_j +\frac{1}{2}}} ^2  F(r) ^2 + C_3 ^2 {Y_{-k} ^{m_j-\frac{1}{2}}}^2 G(r)^2 + C_4 ^2 {Y_{-k} ^{m_j +\frac{1}{2}}}^2 G(r) ^2 ) d\Omega
\end{equation}

Arriving to the complete solution of the relativistic one electron atom. Now we can focus in other applications of the Pauli Matrices.

\section{Representation of the Pauli matrices algebra.}

In this algebra matrices objects are proposed in the following way. 

$$q= q_0 I + q_1 \sigma_z + q_2 \sigma_y + q_3 \sigma_x + q_4 \sigma_y \sigma_x + q_5 \sigma_x \sigma_z + q_6 \sigma_z \sigma_y + q_7 \sigma_x \sigma_y \sigma_z $$

And we can take this object and work it out in the complex quaternion space. 

\begin{equation}
    q= (q_0-iq_7)e_0 -(iq_1+q_4)e_1 - (iq_2+q_5)e_2 - (iq_3+q_7)e_3
\end{equation}

This expression leads us to obtain the algebra of this objects but working it out in a lineal space where the calculations are limited to understanding the non commutativity rule instead of 2x2 matrices.

With this algebra defined we can also represent the Hodge operator in this space.

\begin{equation}
    -\epsilon= -ie_0
\end{equation}

This way we complete the representation of the P algebra.

\section{Representation of the Dirac Matrices.}

The Dirac matrices are the following.

$$ \begin{pmatrix}
        1 & 0 & 0 & 0 \\
        0 & 1 & 0 &0 \\
        0 & 0 & -1 & 0 \\
        0 & 0 & 0 & -1
    \end{pmatrix} ,
    \begin{pmatrix}
        0 & 0 & 0 & 1 \\
        0 & 0 & -1 &0 \\
        0 & -1 & 0 & 0 \\
        1 & 0 & 0 & 0
    \end{pmatrix},
    \begin{pmatrix}
        0 & 0 & 0 & 1 \\
        0 & 0 & 1 &0 \\
        0 & -1 & 0 & 0 \\
        -1 & 0 & 0 & 0
    \end{pmatrix},
    \begin{pmatrix}
        0 & 0 & 0 & -i \\
        0 & 0 & i &0 \\
        0 & i & 0 & 0 \\
        -i & 0 & 0 & 0
    \end{pmatrix}
    $$
These matrices also have a $2x2$ representation using the Pauli matrices.

$$\begin{pmatrix}
       I & 0 \\
        0 & - I
    \end{pmatrix},
    \begin{pmatrix}
       0 & \sigma_z \\
        -\sigma_z &  0
    \end{pmatrix},
    \begin{pmatrix}
       0 & \sigma_x \\
        -\sigma_x &  0
    \end{pmatrix},
    \begin{pmatrix}
       0 & \sigma_y \\
        -\sigma_y &  0
    \end{pmatrix}
    $$
Now applying the complex quaternion representation we arrive to the following matrices.

\begin{equation}
    \begin{pmatrix}
        1 & 0 & 0 & 0 \\
        0 & 1 & 0 &0 \\
        0 & 0 & -1 & 0 \\
        0 & 0 & 0 & -1
    \end{pmatrix}
    => 
    \begin{pmatrix}
       e_0 & 0 \\
        0 &  e_0
    \end{pmatrix}
\end{equation}

\begin{equation}
    \begin{pmatrix}
        0 & 0 & 0 & 1 \\
        0 & 0 & -1 &0 \\
        0 & -1 & 0 & 0 \\
        1 & 0 & 0 & 0
    \end{pmatrix}
    =>
    \begin{pmatrix}
       0 & -ie_1 \\
        ie_1 &  0
    \end{pmatrix}
\end{equation}

\begin{equation}
    \begin{pmatrix}
        0 & 0 & 0 & 1 \\
        0 & 0 & 1 &0 \\
        0 & -1 & 0 & 0 \\
        -1 & 0 & 0 & 0
    \end{pmatrix}
    =>
    \begin{pmatrix}
       0 & -ie_3 \\
        ie_3 &  0
    \end{pmatrix}
\end{equation}

\begin{equation}
    \begin{pmatrix}
        0 & 0 & 0 & -i \\
        0 & 0 & i &0 \\
        0 & i & 0 & 0 \\
        -i & 0 & 0 & 0
    \end{pmatrix}
    =>
    \begin{pmatrix}
       0 & -ie_2 \\
        ie_2 &  0
    \end{pmatrix}
\end{equation}

These matrices can be found in the the Work of Maurits Silvis and they are used to formulate the Dirac equation in a biquaternionic form. (Silvis, M. 2010).

\section{Conclusions.}

In this text we obtained a complete complex quaternion or bi-quaternion representation of the spin 1/2. We used this representation to obtain the rotation operator, spinor functions and solve the relativistic one electron problem, arriving to the probability function. 

This work offers the opportunity to continue on with the applications of the representation, for example, a interesting point is applying the quaternionic spin to quantum computing mathematics in hope of obtaining a gaining in time due to the linear form instead of a matrix. Other ways that this work can be carry over in particle physics, mechanical statistics in the representation of the Pauli Matrices and atomic physics. Other possible application related to the relativistic atoms is the solution of the two electron atom or the Helium atom which is a complex problem due to the upgrade of spins and wave functions. While the Dirac matrices representation can be applied to solve problems related to this expressions in the Dirac space or apply it in general relativity.

For this work we've seen that the representation yields the correct results of the rotation operators and the spinor functions while the complexity of the problem depends on the experience using quaternionic multiplication. Personally the calculation was simpler due to the linear nature, specially in the rotation operators which can be exhausting in the matrix form while dealing with trigonometric functions. 
To conclude this work we seen that the quaternionic representation offers a new way to deal with spin $\frac{1}{2}$ problems and solve complicated equations using a linear form and complex quaternionic properties like the squared norm. 

\subsubsection*{Author Contributions}

This work wouldn't be possible whit-out the supervision of Doctor Marco Atonio Perez de la Rosa professor of Mathematics in the UDLAP. His knowledge in Clifford algebras and their applications in physics allowed me to work and obtain the results presented in this text.
Most of the Identities of complex quaternions used in this text were given by him.

\subsubsection*{Acknowledgments}

I would like to acknowledge Doctor Martin Hentchinsky for helping me obtain an idea in where to apply the biquaternion representation.

Also I would like to thank my Family and my Girlfriend for all the support during the development of this work.

\section{References}

Ávila, R., Abreu, R., and Bory, J. (2016). Álgebra Cuaterniónica y algunas aplicaciones. Con elementos de historia de la ciencia. (1.a ed., Vol. 1). Taller Abierto S.C.L.
 
Conte, E. (2000). Biquaternion quantum mechanics.

Dietterich Labs. (2018, 15 June). How to solve the DIRAC equation for the hydrogen atom | Relativistic Quantum Mechanics [Video]. YouTube. 

Kravchenko, V. V., and Shapiro, M. (1996). Integral representations for spatial models of mathematical physics (Vol. 351). CRC Press.

Pérez-de la Rosa, M. A. (2014). On the Hilbert formulas on the unit sphere for the time-harmonic relativistic Dirac bispinors theory. Journal of Mathematical Analysis and Applications, 416(2), 575-596.

Morita, K. (1983). Quaternionic formulation of Dirac theory in special and general relativity. Progress of theoretical physics, 70(6), 1648-1665.

Sakurai, J. J., $\&$ Commins, E. D. (1995). Modern quantum mechanics, revised edition.

Silvis, M. H. (2010). A quaternion formulation of the Dirac equation. Centre for Theoretical Physics, University of Groningen.

Ward, J.P. Quaternions and Caley Number. Algebra and applications. Dordecht, Boston, London: Kluwer Academic Pubisher ; 1997.

\end{document}